\documentclass[prl,showpacs,twocolumn,superscriptaddress]{revtex4}
\usepackage{amssymb}
\usepackage{amsmath}
\usepackage{graphicx}

\usepackage{color}
\definecolor{blue}{rgb}{0,0,1}
\definecolor{grey}{rgb}{0.6,0.6,0.6}

\def    \bse{\begin{subequations}}
\def    \ese{\end{subequations}}

\newcommand{\ACcomment}[1]{}

\def \be{\begin{equation}}
\def \ee{\end{equation}}
\def \bew{\begin{widetext}\begin{equation}}
\def \eew{\end{equation}\end{widetext}}
\def \bmlett{\begin{mathletters}}
\def \emlett{\end{mathletters}}

\def \ra{\rightarrow}

\def \hrho{\hat{\rho}}

\def \hu{\hat{u}}

\def \ha{\hat{a}}
\def \hb{\hat{b}}
\def \hc{\hat{c}}
\def \hd{\hat{d}}
\def \hH{\hat{H}}
\def \hd{\hat{d}}

\def \omegam{\omega_M}

\def \hcm{\hat{c}}

\def \omegam{\omega_{\rm M}}

\newcommand{\Exp}[1]{\mathrm{e}^{#1}}
\newcommand{\moy}[1]{\langle#1\rangle}
\newcommand{\ket}[1]{|#1\rangle}

\def \hrho{\hat{\rho}}

\begin{document}

\title{Nonlinear interaction effects in a strongly driven optomechanical cavity}

%
%
%

\author{Marc-Antoine Lemonde}
\address{Department of Physics, McGill University, Montr\'eal, QC Canada H3A 2T8}

\author{Nicolas Didier}
\address{Department of Physics, McGill University,  Montr\'eal, QC Canada H3A 2T8}
\address{D\'epartement de Physique, Universit\'e de Sherbrooke,  Sherbrooke, QC Canada J1K 2R1}

\author{Aashish~A.~Clerk}
\address{Department of Physics, McGill University, Montr\'eal, QC Canada H3A 2T8}

\date{Apr. 15, 2013}

\begin{abstract}
We consider how nonlinear interaction effects can manifest themselves and even be enhanced in a strongly driven
optomechanical system.  Using a Keldysh Green's function approach, we calculate modifications to the cavity density of states due to both
linear and nonlinear optomechanical interactions, showing that strong modifications can arise even for a weak nonlinear 
interaction.  We show how this quantity can be directly probed in an optomechanically-induced transparency type experiment.
We also show how the enhanced interaction can lead to nonclassical behaviour, as evidenced by the behaviour of $g_2$ correlation functions.
\end{abstract}


\maketitle



\textit{Introduction-- }
The field of cavity optomechanics involves understanding and exploiting the quantum interaction between a mechanical resonator and photons in a driven electromagnetic cavity.  
It holds immense promise for both fundamental studies of large-scale quantum phenomena as well as applications to quantum information processing and
ultra-sensitive detection, and has seen remarkable progress in the past five years.  Highlights include the use of radiation pressure forces to cool a mechanical resonator to close to its motional
ground state \cite{Teufel2011b,Painter2011b} and experiments where the mechanical motion causes squeezing of the light leaving the cavity \cite{StamperKurn2012,Painter2013}.

As remarkable as this progress has been, it has relied on strongly driving the optomechanical cavity to enhance the basic dispersive coupling between photons and mechanical position.  While the resulting interaction can be made larger than even the dissipative rates in the system \cite{Aspelmeyer2009,Teufel2011,Kippenberg2012}, it is purely bilinear in photon and phonon operators.  As a result, it cannot convert Gaussian state inputs into non-classical states or give rise to true photon-photon interactions.  Recent theoretical work has addressed physics of the nonlinear interaction in weakly driven systems \cite{Rabl2011,Nunnenkamp2011}.  Unfortunately, one finds that effects are suppressed by the small parameter $g / \omega_M$.  

In this paper, we now consider nonlinear interaction effects in an optomechanical system that (unlike Refs.~\cite{Rabl2011,Nunnenkamp2011}) is also subject to a strong laser drive; we consider effects of this driving beyond simple linear-response.  We find somewhat surprisingly that one can use the strong drive to enhance the underlying single-photon interaction.  Using non-equilibrium many-body perturbation theory (based on the Keldysh technique (see, e.g., \cite{Kamenev09})), we calculate how these effects modify the cavity density of states, and hence the cavity's response to an additional weak probe laser.  This response is exactly the quantity measured in so-called optomechanically-induced transparency (OMIT) experiments \cite{Agarwal2010,Kippenberg2010b,SafaviNaeini2011b,Teufel2011}.  We find striking modifications of the OMIT spectrum, effects which can be attributed to the nonlinear interaction causing a hybridization between one and two polariton states (with the polaritons being joint mechanical-photonic excitations).  We also find the possibility of enhanced polariton-polariton interactions, which lead in turn to non-classical correlations (as measured by a $g_2$ correlation function).
    
\ACcomment{Mention:  large driving good, as it causes $\kappa$ to decrease in microwave-cavity systems}

\textit{System--}
The standard Hamiltonian of a driven optomechanical cavity is $\hH = \hH_0 + \hH_{\rm diss}$, with ($\hbar = 1$)
\begin{equation}
	\hH_0=	\omega_{\rm C} \ha^\dag \ha + \omegam  \hb^\dag \hb  + 
			g\left(  \hb^{\dag }+\hb \right)  \ha^\dag \ha + (\sqrt{\kappa} \bar{a}_{\rm in}(t) \ha^\dagger + h.c.). 
\end{equation}
Here $\ha$ is the cavity mode (frequency $\omega _{\rm C}$, damping rate $\kappa$), 
$\hb$ is the mechanical mode (frequency $\omegam$, damping rate $\gamma$), and $g$ is the 
optomechanical coupling.  $\hH_{\rm diss}$ describes dissipation of photons and phonons by independent baths;
$\bar{a}_{\rm in}(t)$ is the amplitude of the drive laser.

We consider the standard case of a continuous-wave drive (i.e.~$\bar{a}_{\rm in}(t) \propto e^{-i \omega_{\rm L} t}$), and work in a rotating
frame at the laser frequency $\omega_{\rm L}$.  We further make a displacement transformation, writing $\ha = e^{-i\omega_{\rm L} t} \left( \bar{a} + \hd \right)$,
where $\bar{a}$ is the classical cavity amplitude induced by the laser drive.  Letting $\Delta = \omega_{\rm L}-\omega_{\rm C}$,
the coherent Hamiltonian now takes the form $\hH_0 = \hH_{1} + \hH_{2}$ with 
\begin{eqnarray}
		\hH_{1} &=&					
			-\Delta \hd^{\dagger} \hd + \omegam \hb^{\dagger} \hb + G (\hd+\hd^{\dagger})(\hb+\hb^{\dagger}),
							\label{eq:H0}		\\
		\hH_{2} &=&					
			g \hd^{\dagger} \hd (\hb+\hb^{\dagger}).
							\label{eq:Hint}
\end{eqnarray}
$G = g \bar{a}$ is the drive-enhanced many-photon optomechanical coupling;  we set $g, \bar{a} > 0$ without loss of generality.  

The most studied regime of optomechanics is where $\bar{a} \gg 1$ and $g \ll \kappa, \omegam$.  It is then standard to neglect the effects of $\hH_2$.  In the absence of any driving, a simple perturbative estimate suggests that the effects of $\hH_2$ enter as $g^2 / \omegam$, where the factor of $\omegam$ corresponds to a virtual state with one extra (or one less) phonon.  This conclusion can be made more precise by exactly solving the coherent, undriven system using a polaron transformation \cite{Rabl2011,Nunnenkamp2011}.  Thus, in this standard regime, one can ignore $\hH_2$, leaving only $\hH_1$, which is easily  diagonalized as $\hH_1 = \sum_{\sigma = \pm} E_\sigma \hc^\dag_{\sigma} \hc_{\sigma}$.  Here $\hc_{+,-}$ describe the two normal modes of the system.  As these modes have both photon and phonon components, we refer to them as polaritons in what follows.  Their energies are:
\begin{align}
		E_{\pm}	&=
			\frac{1}{\sqrt{2}} \left( \Delta^2+\omegam^2 \pm \sqrt{ \left(\Delta^2-\omegam^2 \right)^2-16G^2\Delta\omegam}   \right)^{1/2}.
			\label{eq:PolaritonE}
\end{align}
For $\Delta \simeq -\omega_M$ and $G \geq \kappa, \gamma$, the polariton energy splitting can be resolved experimentally
\cite{Aspelmeyer2009,Teufel2011,Kippenberg2012}.

\ACcomment{Is this the write place to say something about the effective damping and temperature of the polaritons?  Seems so...}



\textit{Polariton interactions-- }
Unlike previous work, we wish to retain the effects of the nonlinear interaction $\hH_2$, but also consider the effects of a large drive (and hence a large many-photon coupling $G$). 
To proceed, we will treat the effects of $\hH_2$ in perturbation theory.  We use a Keldysh Green's function (GF) approach which is able to describe the non-equilibrium nature of the system. 
The linear Hamiltonian $\hH_1$ along with the dissipative terms in $\hH_{\rm diss}$ define the free GFs of the system, which describe the propagation of polaritons in the presence of dissipation.    
Written in the polariton basis, the nonlinear interaction $\hH_2$ gives rise to number-non-conserving interactions,
\begin{eqnarray}
	\hH_2 = \sum_{\sigma,\sigma',\sigma''}  \left( g^A_{\sigma \sigma' \sigma''} \hc^\dag_\sigma \hc^\dag_{\sigma'} \hc^\dag_{\sigma''} + 
		g^B_{\sigma \sigma' \sigma''} \hc^\dag_\sigma \hc^\dag_{\sigma'} \hc_{\sigma''} + h.c. \right),
	\label{eq:HintPolariton}
\end{eqnarray}
where the coefficients $g^{A/B}_{\sigma \sigma' \sigma''} \propto g$ \cite{EPAPS}.  Note normal ordering $\hH_2$ in terms of polariton operators introduces 
small quadratic and linear terms which modify the diagonalized form of $\hH_1$ (see EPAPS for details \cite{EPAPS}).

We start by considering how single-particle properties are modified by the nonlinear interactions; such properties can be directly probed by weakly driving the cavity with a second probe laser (i.e.~an OMIT experiment \cite{Agarwal2010,Kippenberg2010b,SafaviNaeini2011b}) or by measuring the mechanical force susceptibility.  Understanding these properties amounts to calculating the self-energy $\Sigma[\omega]$ of both the polaritons due to $\hH_2$.  We have calculated all self-energy processes to second order in $g$.  Our approach captures both the modification of spectral properties due to the interaction (i.e.~the modification of the cavity and mechanical density of states), as well as modifications of the occupancies of the mechanics and cavity.   While our approach is general, we will focus on the most interesting case of a high mechanical quality factor $\gamma \ll \omegam$, a cavity in the resolved sideband regime $\omegam > \kappa$, and a strong cavity drive, $G \gtrsim \kappa$.  \ACcomment{Actually, we also discuss the regime of a weak drive, don't we?}

\ACcomment{Hmm, we could write the Dyson equation in the case $G >> \kappa$, where the bare polariton GFs are diagonal in polariton number}

Our full second-order self-energy calculation finds that for most choices of parameters, the polariton self-energies scale as $g^2 / \omegam$ and thus have a negligible effect for the 
typical case where $g \ll \omegam$.  However, effects are much more pronounced if one adjusts parameters so that $E_+ = 2 E_-$.  This condition makes the term in $\hH_2$ which scatters a $+$ polariton into 
two $-$ polaritons (and vice-versa) resonant.  It can be achieved for any laser detuning $\Delta$ in the range $(-2 \omegam, -\omegam/2$) by tuning the amplitude $\bar{a}_{\rm in}$ of the driving laser so that the  many-photon optomechanical coupling $G = G_{\rm res}$, where
\begin{equation}
		G_{\rm res}[\Delta]  \equiv  \sqrt{4\Delta^4-17\Delta^2\omegam^2+4\omegam^4}  \, \,  /  \left(10\sqrt{\Delta\omegam} \right).
	\label{eq:Gres}
\end{equation}

In this regime, the dominant physics is well described by the approximation $\hH_0 \simeq \hH_{\rm eff}$ with
\begin{eqnarray}
\label{eq:Heff}
		\hat H_{\rm eff} & = &
		\sum_{\sigma = \pm} E_{\sigma}  \hc_\sigma^{\dagger} \hc_{\sigma}  + \tilde g \left( \hat c_+^{\dagger} \hat c_- \hat c_- + h.c. \right) + \hH_{\rm NR} ,
			\label{eq:Heff}\\
		\hH_{NR} & = &  \sum_{\sigma = \pm} \left(  \delta_{\sigma}  \hc_\sigma^{\dagger} \hc_{\sigma} + 
		\sum_{\sigma' = \pm} U_{\sigma \sigma'} \hc^\dag_{\sigma} \hc^\dag_{\sigma'} \hc_{\sigma'} \hc_{\sigma} \right).
		\end{eqnarray}
The second term in $\hH_{\rm eff}$ corresponds to making a rotating-wave approximation on the nonlinear interaction $\hH_2$ in Eq.~(\ref{eq:HintPolariton}), retaining only the resonant process; 
$\tilde{g} = g^B_{--+} \propto g$ is the
corresponding interaction strength (see inset of Fig.~\ref{fig:OMIT} to see how $\tilde{g}$ varies with $\Delta$).  The terms in $\hH_{NR}$ describe the small (i.e.~$\propto g^2 / \omegam$) residual effects of the non-resonant interaction terms in Eq.~(\ref{eq:HintPolariton}); we treat them via straightforward second-order perturbation theory (i.e.~a Schrieffer-Wolff transformation).  They play no role in the extreme good-cavity limit $\omegam \gg \kappa$ \cite{EPAPS}.



\begin{figure}[t]
	\begin{center}
	\includegraphics[width= 0.95\columnwidth]{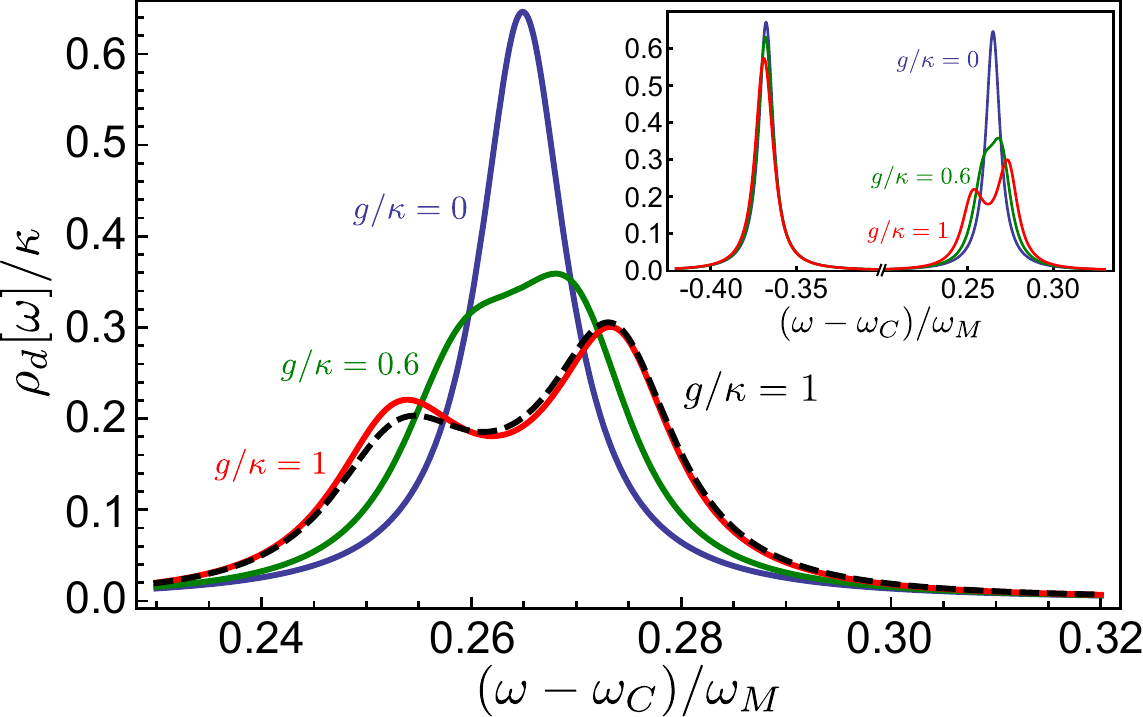}
	\end{center}
	\vspace{-0.5cm} 
	\caption {Main: $+$ polariton resonance in the cavity density of states, 
	for various values of the nonlinear interaction strength $g$ (as indicated), as obtained from Eq.~(\ref{eq:GRplus}) (with the inclusion
	of energy shifts from $\hH_{NR}$ \cite{EPAPS}).  For all plots, the laser drive is at the red sideband
	$\Delta = - \omegam$, and $G = 0.3 \omegam$ to ensure the resonance condition $E_+ = 2 E_-$; we also take
	$\omegam / \kappa = 50$, $T=0$ and $\gamma = 10^{-4} \kappa$.  The peak splitting signals the hybridization of a $+$ polariton with two $-$ polaritons. 
	The dashed curve is the result of a master-equation simulation for $g = \kappa$ \cite{EPAPS}.
	 Inset: Full density of states, same parameters, showing the asymmetry between $+$ and $-$ polariton resonances.}
\label{fig:CavitDOS}
\end{figure}


\textit{Green functions for resonant nonlinear interactions--}
Focusing on the resonant interaction regime defined by Eq.~(\ref{eq:Gres}), and using the simplified Hamiltonian in Eq.~(\ref{eq:Heff}), we obtain simple expressions for the retarded GFs of the 
system.  The retarded photon GF in the displaced, rotating frame is defined as
\begin{equation}\label{eq:GR22}
		G^R_{dd}[\omega] =  -i  \int_{-\infty}^{\infty}  dt \theta(t)
			\left \langle [ \hd(t), \hd^{\dagger}(0) ]  \right \rangle e^{i\omega t},
\end{equation}
with similar definitions for the polariton retarded GF $G^R_{\sigma \sigma}[\omega]$ ($\sigma = \pm$).  As usual, 
$\rho_d[\omega] = - \textrm{Im } G^R_{dd}[\omega] / \pi$ describes the cavity density of states; 
$G^R_{dd}[\omega] $ also determines the reflection coefficient in an OMIT experiment
(see Fig.~\ref{fig:OMIT}).  A standard linear response calculation \cite{EPAPS} shows that the elastic OMIT reflection coefficient  is given by
$r[\omega_{\rm pr}]  =   1 - i  \kappa_{\textrm{cp}} G^R_{dd}[\omega_{\rm pr}]$,
where $\omega_{\rm pr}$ is the frequency of the weak probe beam, and $\kappa_{\textrm{cp}}$ is the contribution to the total cavity $\kappa$ from the coupling to the drive port.

In the limit of interest where $\kappa \ll E_\sigma$, there are no off-diagonal polariton GFs or self-energies \cite{EPAPS}.  As a result, $G^R_{dd}[\omega]$ 
will be given as 
$G^R_{dd}[\omega] = \sum_{\sigma} \left( C_{\sigma} G^R_{\sigma \sigma}[\omega] + D_{\sigma} \left[G^R_{\sigma \sigma}[-\omega] \right]^* \right)$, 
where the change-of-basis coefficients $C_{\sigma}, D_{\sigma} $
are given in \cite{EPAPS}.  The Dyson equations for the polariton retarded GFs are
\begin{eqnarray}
	G^R_{\sigma \sigma}[\omega] & = &
		\left(  \omega - E_{\sigma} + i \kappa_{\sigma}/2 - \Sigma^R_{\sigma \sigma}[\omega] \right)^{-1},
\end{eqnarray}
where $\kappa_{\sigma}$  is the effective damping rate of the $\sigma$ polariton \cite{EPAPS}.  Using the effective Hamiltonian in Eq.~(\ref{eq:Heff}), a standard Keldysh calculation yields that to second order in $g$, the polariton self-energies take the simple forms:
\ACcomment{Should we include the non-resonant shifts here?}
\bse
\label{eqs:Sigmas}
\begin{eqnarray}
	\Sigma_{++}[\omega] & = & \frac{ 2 \tilde{g}^2 (1 + 2 \bar{n}_{-} ) }{ \omega - 2 E_- + i \kappa_+} ,
		\label{eq:Sigma+}\\
	\Sigma_{--}[\omega] & = & \frac{ 4 \tilde{g}^2 (\bar{n}_{-} - \bar{n}_{+}   ) }{ \omega - (E_+ - E_-)  + i (\kappa_+ + \kappa_-)/2} .
		\label{eq:Sigma-}
\end{eqnarray}
\ese
Here, $\bar{n}_{\sigma}$ is the effective thermal occupancy of the $\sigma$ polariton \cite{EPAPS}; for $\tilde{g}=0$, we have $\langle \hc^\dag_{\sigma} \hc_{\sigma} \rangle = \bar{n}_{\sigma}$.  
We have taken the limit $g / \omegam \ra 0$, and hence neglected the effects of the non-resonant terms $\hH_{\rm NR}$ in Eqs.~(\ref{eqs:Sigmas}); the explicit corrections due to these terms are given in the supplemental information \cite{EPAPS}.

Eqs.~(\ref{eqs:Sigmas}) are central results of this work.  Eq.~(\ref{eq:Sigma+}) describes the fact that a single 
$+$ polariton can resonantly turn into two $-$ polaritons, and describes the hybridization between these states that occurs for large enough $g$.  
To see this explicitly, we consider the case of exact resonance (i.e.~$E_+ = 2 E_-$) and write
\begin{eqnarray}\label{eq:GRplus}
		G^R_{+}[\omega] & = & 
			\frac{1}{2} \sum_{\eta = \pm} 
			\frac{1 - i \eta \frac{2\tilde{\kappa}_--\tilde{\kappa}_+}{4\delta_+}}{\omega-E_++i\frac{2\tilde{\kappa}_-+\tilde{\kappa}_+}{4} + \eta \delta_+}, \\
		\delta_{+}	& = &
			\sqrt{2\tilde g^2 \left( 1+2 \bar{n}_- \right)-\left( 2\tilde{\kappa}_- - \tilde{\kappa}_+ \right)^2/16 }.\label{second}
\end{eqnarray}
For $\tilde{g} \gtrsim \kappa$, we see that the $+$ polariton GF has two poles, corresponding to the new hybridized eigenstates. We stress that these eigenstates do not correspond to a fixed 
excitation number.  Note that unlike the undriven system \cite{Rabl2011,Nunnenkamp2011}, the effects of the nonlinear interaction can be significant even if $g \ll \omegam$.
Also note that the resonant coupling between $|+\rangle $ and $|-- \rangle$ states is enhanced at finite temperature by a standard stimulated emission factor
$(1 + 2 \bar{n}_{-})$.   The form of this GF and self-energy are reminiscent to the photon GF for ordinary OMIT, where a photon can resonantly turn into a phonon \cite{Agarwal2010}; however, that effect does not involve any temperature-dependent enhancement.
$\Sigma_{--}$ in Eq.~(\ref{eq:Sigma-}) describes a process where the propagating $-$ polariton of interest interacts with an already-present $-$ polariton to turn into a $+$.  As this process requires an existing density of polaritons, it is strongly suppressed at low temperatures.

We note that it is possible to use resonance to enhance the nonlinear optomechanical interaction without strong driving, if one instead considers a system where two cavity modes interact with a single mechanical resonator \cite{Ludwig2012,Rabl2012,Bennett2013}.  Our approach has the benefit of only requiring a single cavity mode; further, for drive detunings near $\Delta = - \omegam$, it also
has a natural resistance against mechanical heating, as mechanical contribution to the polariton temperature 
scales as $\gamma \bar{n}_{\rm th} / \kappa$, where $\bar{n}_{\rm th}$ is the mechanical thermal occupancy.  While a low temperature
is not essential for the density-of-states effects described above, it is essential for the correlation effects discussed below.
Finally, for superconducting
microwave cavities, the cavity linewidth $\kappa$ has a strong contribution from two-level fluctuators, and thus improves if one strongly drives the cavity (as in our scheme).

\textit{Red-sideband drive-- } For a detuning $\Delta = -\omegam$, the polariton resonance occurs when $G = 0.3 \omegam$.  For this detuning, 
both polaritons are almost equal mixtures of photon and phonon operators.  One finds $\kappa_{\sigma} = (\kappa + \gamma) / 2$, and 
that the resonant interaction strength $\tilde{g} \simeq -0.37 g$.
Because $\hH_1$ does not conserve the number of photons and phonons, the 
polaritons are not eigenstates of $\hd^\dag \hd + \hb^\dag \hb$; as a result, even at zero temperature, 
the effective thermal occupancies scale as $\bar{n}_\sigma \propto (G/\omegam)^2 \ll 1$ \cite{EPAPS}.  
 The inset of Fig.~\ref{fig:CavitDOS} shows the evolution of the cavity density of states for these parameters as $g$
 is increased from zero.  For $g=0$, one sees two symmetric peaks corresponding to the two polaritons, i.e.~the well known normal-mode splitting \cite{Marquardt07,Kippenberg2008b}.  As $g$ increases, these peaks develop a marked asymmetry.  For $g \sim \kappa$,  a clear splitting of the $+$ peak occurs, corresponding to the resonant hybridization of one and two polariton states.  Fig.~\ref{fig:CavitDOS} also shows results of a numerical (but non-perturbative) master-equation simulation \cite{EPAPS}, showing our analytic approach is reliable even for moderately
 strong $g$.
  
 \textit{Large-detuned drives-- }The resonant-polariton interaction is also interesting for drives far from the red-sideband, where the value of $G_{\rm res} \ll \omegam$.  For a laser detuning near the minimum possible value $\Delta = - 2 \omegam$ at which resonance is possible (and setting $G = G_{\rm res}$), the polaritons are each either almost entirely phonon or photon, implying 
 a very small value of $\tilde{g} \propto g G / \omegam$.  However, as the $-$ polariton is now almost purely phononic, its small damping rate and potentially 
 large thermal occupancy enhances the self-energy in Eq.~(\ref{eq:Sigma+}) 
 (i.e.~$\kappa_- \simeq \gamma$, and $\bar{n}_{-}$ corresponds to the mechanical thermal occupancy).  We can quantify these effects by considering the value of $\rho_d[\omega = E_+]$, which will be suppressed by the hybdriziation physics described here.  One finds:
 \begin{eqnarray}
 	\rho_d[E_+] =
		\frac{2 / \pi }{\kappa_+} \frac{1}{1 +  C_{\rm eff} },
		\hspace{0.3cm}  
 	C_{\rm eff} = \frac{ 4 \tilde{g}^2 (1 + 2 \bar{n}_{-} )}{  \kappa_+ \kappa_-}.
	\label{eq:Ceff}
\end{eqnarray}
\ACcomment{Check factors of 2!}  For a large detuning, the effective cooperativity scales as $C_{\rm eff} \propto C (g / \omegam)^2$, where $C = 4 G^2 / \kappa \gamma$ is the 
standard many-photon coupling cooperativity.  Thus, in the large-detuned regime, resonant polaritons interactions allow one to amplify the effects of the nonlinear interaction by a factor $C$.   Fig.~\ref{fig:OMIT} shows the evolution of the OMIT reflection coefficient (which reflects the structure in $\rho_d[\omega]$) as the detuning $\Delta$ is varied, while keeping $G$ tuned to the resonant value $G_{\rm res}(\Delta)$.


\begin{figure}[t]
	\begin{center}
	\includegraphics[width= 0.95\columnwidth]{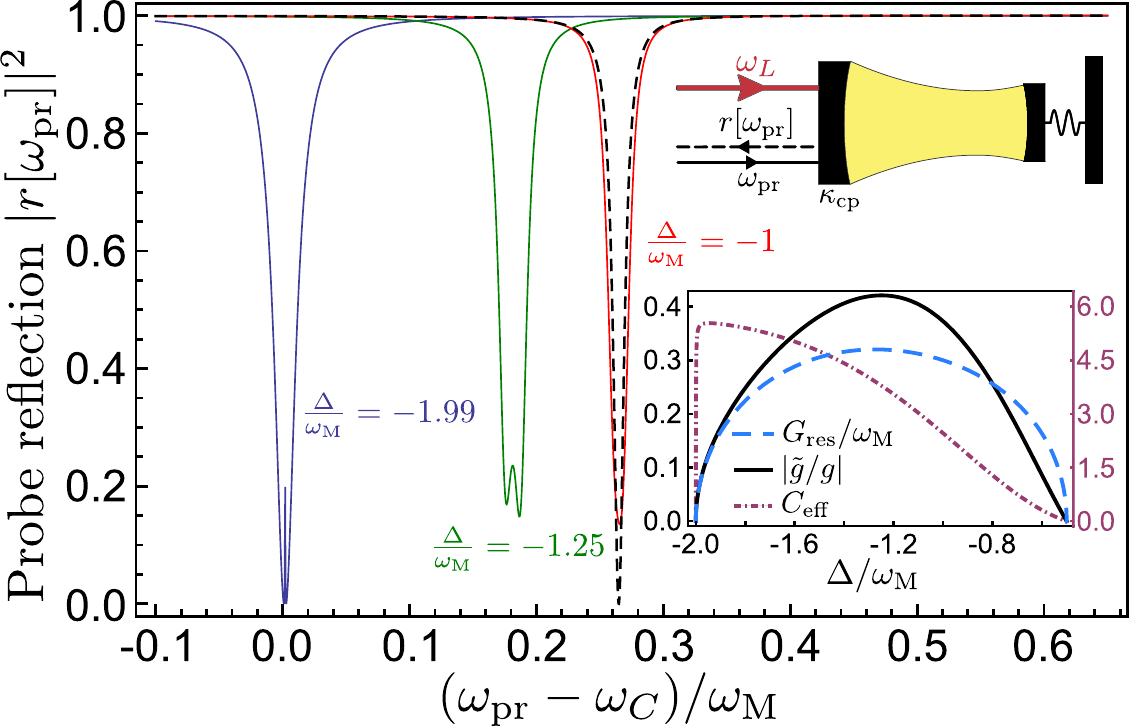}
	\end{center}
	\vspace{-0.5cm} 
	\caption {Reflection coefficient for a weak probe beam incident at a frequency $\omega_{\rm pr}$ (defined in the lab frame), 
	as measured in an OMIT experiment (upper inset).  We take a one sided cavity
	with $\kappa_{\rm cp} / \kappa = 0.5$,  $g = 0.5 \kappa$.  For each curve, $\Delta$ is labelled, and  
	$G= G_{\rm res}[\Delta]$.   
	Remaining parameters are the same as Fig. 1.  Lower inset: behaviour of $g$, $G_{\rm res}$ and $C_{\rm eff}$
	as a function of detuning $\Delta$ of the main laser drive.}
\label{fig:OMIT}
\end{figure}

\begin{figure}[t]
	\begin{center}
	\includegraphics[width= 0.90\columnwidth]{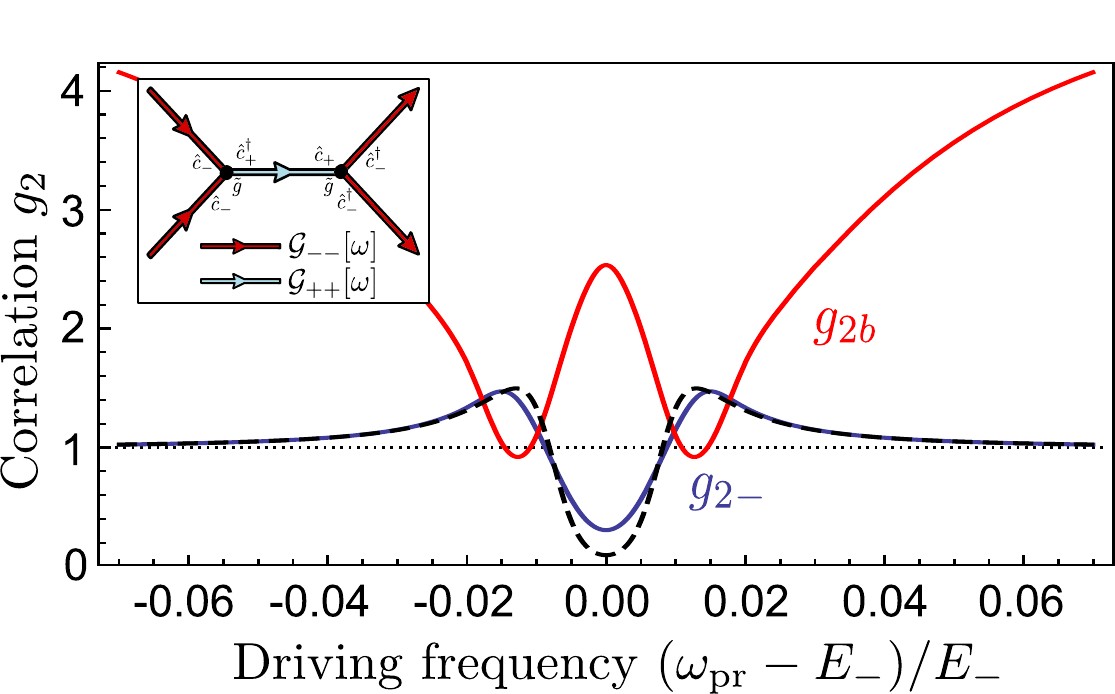}
	\end{center}
	\vspace{-0.5cm} 
	\caption {Inset: Resonant interaction between $-$ polaritons.  Main:  
	Numerically-calculated 
	$g_2$ correlation function for $-$ polaritons ($g_{2-}$) and phonons ($g_{2b}$), in the presence of an additional 
	weak probe laser (frequency $\omega_{\rm pr}$).  Here, $g = \kappa$, $\Delta = -\omegam$, $G = 0.3 \omegam = G_{\rm res}$.
	We have taken $\omegam / \kappa \rightarrow \infty$ to suppress non-resonant interaction effects.  
	The probe amplitude is $\epsilon = 0.2 \kappa$ ($g_{2-}$), $\epsilon = 0.3 \kappa$ ($g_{2b}$).
	Both phonon and polariton $g_2$ functions drop below $1$ due to the interactions, indicating non-classical correlations despite the fact $g / \omegam \simeq 0$.  
	The dashed curve is the result
	of an analytic theory  (see \cite{EPAPS}).}
\label{fig:g2}
\end{figure}

\textit{Induced Kerr interaction-- } 
The nonlinear interaction in the resonant regime defined by Eqs.~(\ref{eq:Gres})-(\ref{eq:Heff})
leads to a strongly enhanced two-particle interaction between $-$ polaritons, mediated by the exchange of a $+$ polariton (Fig.~\ref{fig:g2}).  
In a weakly-driven optomechanical system, Eq.~(\ref{eq:Hint}) implies that phonons can mediate an effective 
photon-photon interaction; however, as the virtual phonon is off-resonance, this interaction $U \propto g^2 / \omegam$.  In contrast, the resonance condition
$E_+ = 2 E_-$ yields an induced interaction $U_{\rm res} \propto \tilde{g}^2 / \kappa$, an enhancement by a large factor $\propto \omegam / \kappa$.  

To assess the effects of the polariton-polariton interactions,   
we weakly drive our system with a second probe tone, and consider the $g_2$ correlation functions  
$g_{2u} = \langle \hu^\dag \hu^\dag \hu \hu \rangle / \langle \hu^\dag \hu \rangle^2$, where $u = b,d, c_+, c_-$.  $g_{2u}$ is a measure of interaction induced correlations;
$g_{2} \leq 1$ signifies non-classical correlation.  Given the strong interaction
experienced by $-$ polaritons when the resonance condition $E_+ = 2 E_-$is achieved, we expect that if the cavity is driven at the $E-$ resonance, $g_{2-}$ will drop below $1$.
This is indeed the result of  a numerical, master-equation based calculation (see Fig.~\ref{fig:g2} and \cite{EPAPS}).  An analytic calculation based on a reduced state-space (similar to that in Ref.~\cite{Bennett2013}) reproduces these results.  For a weak probe drive at the $E-$ frequency, it yields \cite{EPAPS}:
\begin{eqnarray}
	g_{2-}  = \frac{1}{1+ 4 \tilde{g}^2 / \kappa_-^2}.
\end{eqnarray}
One also finds non-classical correlations for photons and phonons.  Shown in Fig.~\ref{fig:g2} is the phonon $g_2$ function $g_{2b}$ 
(for same parameters); it clearly drops below $1$.  The double-peak structure of this curve is the result of the drive inducing correlations between $-$ and $+$ polaritons; it also occurs in 
the behaviour of $\langle \hb^\dag \hb \rangle$ (see EPAPS for more details \cite{EPAPS}). 
%
%


\textit{Conclusions-- }
We have presented a systematic approach for describing nonlinear interaction effects in a driven optomechanical system, identifying
a regime where a resonance enhances interactions between polaritons.  We have discussed how this would manifest itself in a OMIT-style experiment, as well as in $g_2$ correlation functions.  The polariton interactions we describe could be extremely interesting when now considered in lattice systems, or when considering the propagation of
pulses. 

We thank W. Chen and A. Nunnenkamp for useful discussions.  This work was supported by CIFAR, NSERC and the DARPA ORCHID program under a grant from AFOSR.
\textit{Note added-- }During the preparation of this paper, we became aware of a related work by B\o rkje, Nunnenkamp, Teufel and Girvin.


\bibliographystyle{apsrev}



\begin{widetext}

\newpage

\renewcommand{\theequation}{S\arabic{equation}}

\setcounter{equation}{0}

\section{Supplemental information}


\section{Langevin equations for linearized optomechanical system in the polariton basis}

We start by diagonalizing the linearized optomechanical Hamiltonian given in Eq.~(2) of the main text, working as always in a displaced interaction picture set by the laser drive on the cavity.
Introducing
\begin{eqnarray}
        \vec{X} &=&\left[
        \begin{array}{cccc}
            \hat{b} & \hat{d} & \hat{b}^{\dagger } & \hat{d}^{\dagger }
        \end{array}
        \right] ^{T},  \,\,\,\,\,\,
        \vec{Y} = \left[
        \begin{array}{cccc}
            \hat{c}_- & \hat{c}_+ & \hat{c}_-^{\dagger } & \hat{c}_+^{\dagger }
        \end{array}
        \right] ^{T},
\end{eqnarray}
the diagonalization can be expressed in terms of a $4 \times 4$ real matrix $\mathbf{U}$:    
\begin{equation}
	\vec{Y} = \mathbf{U} \cdot \vec{X}.
	\label{eq:UTransform}
\end{equation}
$\mathbf{U}$ can be found by standard means, though its general form is both cumbersome and unenlightening; we define $\mathbf{V} = \mathbf{U}^{-1}$.  It is slightly less unwieldy in the special (but relevant) case of a drive at the red-detuned mechanical sideband $\Delta = -\omegam$.  In this case, we have simply:
\begin{equation}
	\hcm_\pm=\frac{1}{\sqrt{8\omega_ME_\pm}}\left[ (E_\pm-\omega_M)(\hd^\dag\pm\hb^\dag) + (E_\pm+\omega_M)(\hd\pm\hb) \right],
\end{equation}
while the inverse transformation is defined by:
\begin{eqnarray}
	V_{1j} & = &
		\frac{1}{\sqrt{8 \omegam}}
		\left[
		\begin{array}{cccc}
			-\frac{\omegam + E_-}{\sqrt{E_{-}}},	&	\frac{\omegam + E_+}{\sqrt{E_{+}}},	&
			-\frac{\omegam - E_-}{\sqrt{E_{-}}},	&	\frac{\omegam - E_+}{\sqrt{E_{+}}}	\\
		\end{array}
		\right]	\\
	V_{2j} & = &
		\frac{1}{\sqrt{8 \omegam}}
		\left[
		\begin{array}{cccc}
			\frac{\omegam + E_-}{\sqrt{E_{-}}},	&	\frac{\omegam + E_+}{\sqrt{E_{+}}},	&
			\frac{\omegam - E_-}{\sqrt{E_{-}}},	&	\frac{\omegam - E_+}{\sqrt{E_{+}}}	\\
		\end{array}
		\right]	
\end{eqnarray}

Next, we include the coupling of our system to the cavity and mechanical dissipative baths in the standard way, treating these baths as Markovian over frequencies of interest.  Consider first the coupling to the cavity bath.  Prior to making displacement and interaction-picture transformations, the system-bath coupling has the form (see, e.g., Ref.~\onlinecite{ClerkRMP}):
    \begin{eqnarray}
        \hH_{\kappa} = \sum_j \omega_j \hat{f}^\dagger_j \hat{f}_j, \,\,\,\,         
        \hat{H}_{\kappa,int} &=& -i \sqrt{\frac{\kappa }{2\pi \rho_c }}\sum_{j}\left( \hat{f}_{j} - \hat{f}_{j}^{\dagger }\right) \left( \hat{a}+\hat{a}^{\dagger }\right),
    \end{eqnarray}
where $\hat{f}_{j}$ is the lowering operator for a bath mode, and $\rho_c$ is the density of states of bath modes (which we treat to be constant over frequencies of interest).  In now moving to an interaction picture at the laser frequency, we also transform the bath modes, i.e. $\hat{f}_j \rightarrow \hat{f}_j e^{-i \omega_L t}$. Formally, the interaction picture transformation involves a unitary $\hat{U}(t) = \exp\left[ -i \omega_L t  \left(\ha^\dagger \ha + \sum_j \hat{f}^\dagger_j \hat{f}_j \right) \right]$
and transforms the bath Hamiltonian to $\hH_{\kappa} = \sum_j (\omega_j - \omega_L) \hat{f}^\dagger_j \hat{f}_j$. 
In this interaction picture, the counter-rotating terms in 
$\hat{H}_{\kappa,int}$ will explicitly oscillate at $\pm 2 \omega_L =\pm 2 (\omega_{c} + \Delta)$.  Even if we now write our photon operators $\hat{a}$ in terms of polariton operators, there is no possibility of having these terms becoming resonant, as the cavity frequency is much larger than any other frequency scale in the problem (i.e. $\omega_c \gg | \Delta|, \omegam, E_+, E_-$).  As such, one can safely make a rotating-wave approximation in the photon basis, resulting in a standard system-bath interaction which is stationary in the interaction picture:
 \begin{eqnarray}
       \hat{H}_{\kappa,int} &=& i \sqrt{\frac{\kappa }{2\pi \rho_c }}\sum_{j}\left(  \hat{f}_{j}^{\dagger } \hd  - \hat{d}^\dag \hat{f}_{j}   \right) \nonumber \\
       & = & 	-i \sqrt{\frac{\kappa }{2\pi \rho_c }}\sum_{j} \left[ \hat{f}_{j}^{\dagger } \left( V_{21} \hc_{-} + V_{22} \hc_{+} + V_{23} \hc^\dagger_{-} + V_{24} \hc^\dagger_{+} \right) - h.c. \right] 
       \label{eq:HCavBath}
 \end{eqnarray}%
We have also made the displacement transformation $\ha = \hd + \bar{a}$ as discussed in the main text.  Note that in writing $\hat{d}$ in terms of polariton operators via Eq.~(\ref{eq:UTransform}), we obtain terms of the form $\hat{f}^\dagger_j \hat{c}_{\sigma}^\dagger$, which can cause polariton heating even if the cavity bath is at zero temperature.  Such terms are physical and must be retained.  Formally, In the interaction picture the bath now has negative frequency modes which can make such processes resonant.  In more physical terms, the combination of zero-point bath fluctuations with the cavity driving can excite the polaritons.  This mechanism has been discussed in other contexts under the name ``quantum activation" by various authors \cite{Dykman2006,Wilhelm2007}.
    
We turn now to the mechanical bath, where the basic interaction Hamiltonian can be written in an analogous way:
    \begin{eqnarray}
            \hH_{\gamma} = \sum_j \omega_j \hat{g}^\dagger_j \hat{g}_j, \,\,\,\,         
        \hat{H}_{\gamma,int} &=& -i \sqrt{\frac{\gamma }{2\pi \rho_m }}\sum_{j}\left( \hat{g}_{j} - \hat{g}_{j}^{\dagger }\right) \left( \hat{b}+\hat{b}^{\dagger }\right).
    \end{eqnarray}
Here, $\hat{g}_j$ is a lowering operator for a mechanical bath oscillator, and $\rho_m$ is the density of states of mechanical bath modes (which we also treat to be constant over frequencies of interest).  As there is no direct driving of the mechanical resonator, the analysis here is somewhat simpler.  We first re-write the phonon operator $\hb$ in the polariton basis, and {\it then} make a rotating-wave approximation.  The justification is that counter-rotating terms such as $\hat{g}^\dagger_j \hat{c}^\dagger_{\sigma}$ can never be made resonant; as the mechanical resonator is not driven, there is no quantum activation mechanism involving the mechanical bath.  We thus obtain
   \begin{eqnarray}
        \hat{H}_{\gamma,int}   & \simeq & \sum_{\sigma }\sqrt{\frac{\kappa^m _{\sigma}}{2\pi \rho_m }}%
        \sum_{j}\left( \hat{g}_{j}^{\dagger }\hat{c}_{\sigma }+h.c.\right),
        \label{eq:HMechBath}
    \end{eqnarray}%
where ($\mathbf{V} = \mathbf{U}^{-1}$)
\begin{eqnarray}
	\kappa^M_{-} = \gamma \left( V_{11} + V_{31}\right) ^{2}, \,\,\,\,\,\,\,
	\kappa^M_{+} = \gamma \left( V_{12} + V_{32} \right)^2.
\end{eqnarray}

Having now established the correct system-bath coupling Hamiltonians in the polariton basis we wish to use, we can derive the Heisenberg-Langevin equations
for our system in the standard manner.  
As each bath couples to both $+$ and $-$ polaritons, one obtains off-diagonal damping terms, (e.g.~the mechanical bath produces a force on the $+$ polariton that is proportional to the amplitude
of the $-$ polariton).  Such terms will dynamically couple $+$ and $-$ polaritons, and can be included in our theory in a straightforward manner (i.e.~by including Green functions that
are off-diagonal in the polariton index).  However, in the 
weak-dissipation limit of interest ($\kappa, \gamma \ll E_+, E_-, |E_+ - E_-|$), the mixing effects induced by such terms is strongly suppressed.  As such, 
we will drop off-diagonal damping terms, resulting in the form:
\begin{eqnarray}
        \partial _{t}\hat{c}_{\sigma}(t) & = &-\left( iE_{\sigma}+\frac{\kappa _{\sigma}}{2}\right) \hat{c}_{\sigma}(t)-\sqrt{\kappa _{\sigma}}\hat{\xi}_{\sigma}(t),
        \label{eq:Langevins} 
 \end{eqnarray}
 where the polariton damping rates $\kappa_{\sigma}$ are given by:
 \begin{eqnarray}
 	\kappa_- = \kappa^M_{-} + \kappa \left[  \left( V_{21}\right)^2 - \left( V_{23}\right)^2  \right],
	\,\,\,\,\, 
 	\kappa_+ = \kappa^M_{+} + \kappa \left[  \left(V_{22}\right)^2 - \left( V_{24} \right)^2  \right] .
 \end{eqnarray}
The noise operators $\hat{\xi}_\sigma(t)$ are each linear 
combinations of the input noise emanating from the mechanical and cavity baths.  In the interaction picture we use, we will be sensitive to noise
in the cavity bath at frequencies near $\omega_c$, and noise in the mechanical bath at frequencies near $\omega_M$.  In the limit of interest where the physical temperature $T \ll \hbar \omega_c / k_B$, there will be no thermal noise incident from the cavity bath at the frequencies of interest.  Also, as we focus on regimes where the polariton damping rates are much smaller than their energies, we can treat the noise operators as being white noise (as is standard in input-output theory treatments).  We thus have
 \begin{eqnarray}
        \left\langle \hat{\xi}_{\sigma }^{\dagger }(t) \hat{\xi}_{\sigma }(t^{\prime})\right\rangle 
        		& = & 
			\overline{n}_{\sigma }\delta \left( t-t^{\prime }\right), 
		\,\,\,\,\,\,
        \left\langle \hat{\xi}_{\sigma }(t) \hat{\xi}_{\sigma }^{\dagger }(t^{\prime})\right\rangle 
        		 =  
			\left(1 + \overline{n}_{\sigma } \right) \delta \left( t-t^{\prime }\right),
 \end{eqnarray}
where the effective temperatures of the two noises are given by:
 \begin{eqnarray}
        \overline{n}_{-} & = &\frac{1}{\kappa _{-}}\left( \kappa^M_- n_{B}\left[ E_{-}\right]+\left(V_{23}\right)^{2}\kappa \right) , 
        \,\,\,\,\,\,\,
        \overline{n}_{+}  = \frac{1}{\kappa _{+}}\left( \kappa^M_+ n_{B}\left[ E_{+}\right] + \left(V_{24}\right)^{2}\kappa \right).
 \end{eqnarray}%
$n_{B}\left[ E_{\sigma }\right] $ denotes the Bose-Einstein distribution function evaluated at energy $E_{\sigma }$ and the mechanical bath temperature.

Finally, one also finds that $\hat{\xi}_+$ and $\hat{\xi}_-$ are correlated with one another.  Similar to the situation of off-diagonal damping terms, such noise correlations could easily be included in our theory; however, as they play no role in the regime of interest where $E_{\sigma}, |E_+ - E_-| \gg \kappa, \gamma$, we drop them in what follows.  One also finds that anomalous noise
correlators can be non-zero (e.g.~$\langle \hat{\xi}_\sigma (t) \hat{\xi}_{\sigma'} (0) \rangle$).  Again, while such terms can be retained in our theory, they play no role in the weak-damping regime of interest, and we hence drop them in what follows.

\section{Unperturbed polaritons GFs}

The standard definitions of the three relevant GFs needed in the Keldysh technique are ($\sigma $ stands for $\pm $, but similar definitions hold for $\hat{c}_{\sigma }=\hat{b}$ or $\hat{d}$):
    \begin{eqnarray}
        G_{\sigma \sigma' }^{R}\left[ \omega \right]  &\equiv &
        -i\int_{-\infty}^{\infty} dt\theta(t)\left\langle \left[ \hat{c}_{\sigma }(t),\hat{c}_{\sigma' }^{\dagger }(0)\right] \right\rangle e^{i\omega t},  \label{eq_def_GR} \\
        G_{\sigma \sigma' }^{A}\left[ \omega \right]  &\equiv &
        i\int_{-\infty}^{\infty} dt\theta(-t)\left\langle \left[ \hat{c}_{\sigma }(t),\hat{c}_{\sigma' }^{\dagger }(0)\right] \right\rangle e^{i\omega t},  \label{eq_def_GA} \\
        G_{\sigma \sigma' }^{K}\left[ \omega \right]  &\equiv &
        -i\int_{-\infty}^{\infty} dt\left\langle\left\{ \hat{c}_{\sigma }(t),\hat{c}_{\sigma' }^{\dagger }(0)\right\}\right\rangle e^{i\omega t}.  \label{eq_def_GK}
    \end{eqnarray}%
The retarded and advanced GFs keep track of spectral information, whereas the Keldysh Green function $G^K$ also keeps track of the occupancy of states.  For the linearized 
(non-interacting) theory, the GFs are easily obtained from the Langevin equations in Eq.~(\ref{eq:Langevins}).  These free GFs (which we denote by $\mathcal{G}$) are diagonal (only non-zero for $\sigma = \sigma'$) and given by:
 \begin{eqnarray}
        \mathcal{G}_{\sigma \sigma }^{R}\left[ \omega \right]  &=&\frac{1}{\omega-E_{\sigma }+i\frac{\kappa _{\sigma }}{2}}, \\
        \mathcal{G}_{\sigma \sigma }^{K}\left[ \omega \right]  &=&
        \left( 2\overline{n}_{\sigma }+1\right) \left( \mathcal{G}_{\sigma \sigma }^{R}\left[ \omega \right]-\mathcal{G}_{\sigma \sigma }^{A}\left[ \omega \right] \right),
    \end{eqnarray}
and $\mathcal{G}_{\sigma \sigma }^{A}\left[ \omega \right] =\left[ \mathcal{G}_{\sigma \sigma}^{R}\left[ \omega \right] \right] ^{\ast }$.

Finally, as there are no off-diagonal polariton Green functions, we can use Eq.~(\ref{eq:UTransform}) to write the photon retarded Green function $G^R_{dd}[\omega]$ as:
\begin{eqnarray}
	G^R_{dd}[\omega] = \sum_{\sigma} \left( C_{\sigma} G^R_{\sigma \sigma}[\omega] + D_{\sigma} \left[G^R_{\sigma \sigma}[-\omega] \right]^* \right), 
\end{eqnarray}
where
\begin{eqnarray}
	C_{-}  =  V_{21}^2, \,\,\,  
	C_{+} = V_{22}^2, \,\,\,
	D_{-}  =  V_{23}^2, \,\,\,
	 D_{+} = V_{24}^2.
\end{eqnarray}
As the Green functions remain diagonal in the polariton index even with the nonlinear interaction (in the regime of interest, see below), the above relation also holds for the full Green functions 
(i.e.~including the self-energy associated with $g$).

\section{$\hat{H}_{2}$ in the polariton basis}

Using the change of basis matrix $\mathbf{V} = \mathbf{U}^{-1}$ (c.f. Eq.(\ref{eq:UTransform})), we can re-write the non-linear interaction $\hH_2$ in Eq.~(3) of the main text in the polariton basis via:
\begin{equation}
        \hat{H}_{2}=g\left[ \mathbf{V} \vec{Y}\right] _{4}\left[\mathbf{V} \vec{Y}\right] _{2}\left( \left[ \mathbf{V}\vec{Y}\right] _{1}%
        +\left[ \mathbf{V} \vec{Y}\right] _{3}\right) .
\end{equation}%
Expanding this equation allows one to obtain the interaction coefficients $g^{A,B}_{\sigma \sigma' \sigma''}$ in terms of $g$ and matrix elements of $\mathbf{V}$.  In particular, the 
coefficient $\tilde{g}$ of the resonant interaction process $\hat{c}^\dagger_- \hat{c}^\dagger_- \hat{c}_+$ will be given by:
\begin{equation}\label{eq:gtilde}
	\tilde{g}  =  g (V_{22}V_{21}(V_{11}+V_{13})+V_{23}V_{24}(V_{11}+V_{13})+V_{23}V_{21}(V_{14}+V_{12})).
\end{equation}
As the normal-mode transformation described by $\mathbf{U}$ mixes raising and lowering operators, $\hH_2$ will not be normal-ordered when written in terms
of polariton operators (even though it is normal ordered when written in terms of photon and phonon operators).  Normal-ordering $\hH_2$ in the polariton basis
yields the form:
\begin{equation}
	\hH_2 = : \hH_2 :   +\left( A_{-}\hat{c}_{-}+A_{+}\hat{c}_{+}+h.c.\right)
\end{equation}
where the colons indicate normal-ordering in the polariton basis, and 
the constants $A_{\sigma} \propto g$.  The first term is the normal-ordered polariton interaction written in Eq.~(5) in the main text.
We next make a unitary displacement transformation of the form $\hat{c}_\sigma \rightarrow \bar{c}_{\sigma} + \hat{c}_{\sigma}$,
where the constants $\bar{c}_{\sigma}$ are chosen to eliminate all linear-in-$\hat{c}_{\sigma}$ terms in the Hamiltonian; to leading order in $g$,
$\bar{c}_{\sigma} = - A_\sigma / E_{\sigma}$.  The resulting coherent Hamiltonian has the form:
    \begin{eqnarray}
    	\hH_2 &= &
        : \hat{H}_{2}:   +  
        \left[
        	\begin{array}{cc}
            \hat{c}_{-}^{\dagger } & \hat{c}_{+}^{\dagger }
        \end{array}
        \right]
        \mathbf{Z_1}
       \left[
        \begin{array}{c}
            \hat{c}_{-} \\
            \hat{c}_{+}
        \end{array}
        \right]  
	+
        \left( \left[
        \begin{array}{cc}
            \hat{c}_{-} & \hat{c}_{+}
        \end{array}
        \right] 
        \mathbf{Z_2}
        \left[
        \begin{array}{c}
            \hat{c}_{-} \\
            \hat{c}_{+}
        \end{array}
        \right] +h.c.\right) .  \notag
    \end{eqnarray}
Here $\mathbf{Z_1}, \mathbf{Z_2}$ are
$4 \times 4$ matrices whose entries are all order $g^2 / E_{\sigma} \sim g^2 / \omegam$.  The quadratic terms on the RHS describe corrects to the linear Hamiltonian 
$\hH_1 = \sum_\sigma E_{\sigma} \hat{c}^\dagger_{\sigma} \hat{c}_{\sigma}$ arising from the displacement transformation.  
In principle, one could combine these with the terms in $\hH_1$, re-diagonalize the resulting quadratic Hamiltonian, obtaining 
a new basis of non-interacting polaritons.  However, to leading order in $g^2 / \omegam$, all that is needed is to retain the diagonal elements of $\mathbf{Z_1}$, which simply shift the polariton energies but do not change their wavefunctions.  These energy shifts (denotes $\epsilon_+, \epsilon_-$) are just absorbed into the definition of our free Green functions.  We are then left with a normal-ordered interaction Hamiltonian $\hH_2$ in the polariton basis which can be addressed perturbatively. 
   
%


\section{Perturbative treatment}


\begin{figure}[t]
	\begin{center}
	\includegraphics[width= 0.95\columnwidth]{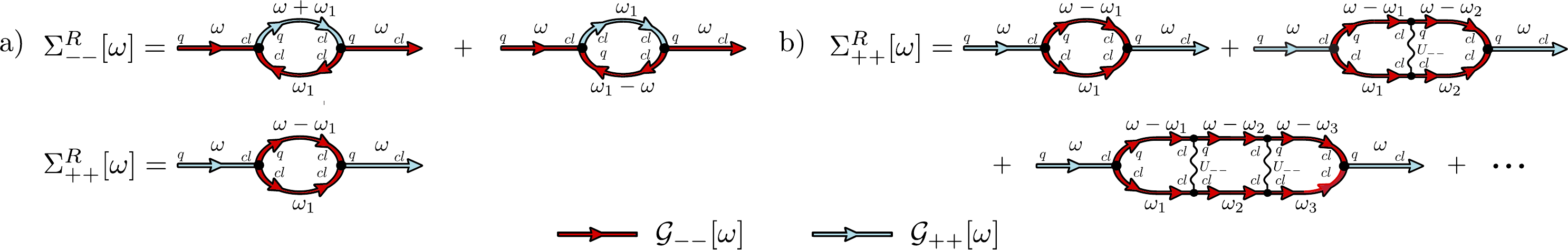}
	\end{center}
	\vspace{-0.5cm} 
	\caption {(a) Feynman diagrams describing the dominant $+$ and $-$ polariton self-energies (to second order in $g$) 
	in the case  when the resonance condition $E_+ = 2 E_-$ is met.  We have shown diagrams for the retarded self-energies; 
	similar diagrams also determine the advanced and Keldysh self-energies.  The structure of propagators in Keldysh space is indicated by writing
	explicit classical ($cl$) and quantum ($q$) indices at the ends of each Green function.
	(b) The non-resonant interaction terms induce effective two-particle interactions between polaritons, as described by $H_{NR}$ in Eq.~(8) of the main text.
	These two-particle interactions modify the resonant self-energies depicted in Fig.~(\ref{fig:SigmaRPol}); shown above are the diagrams
	describing the modification of the $+$ polariton self-energy by the effective interaction $U_{--}$.  The summation of these ladder diagrams results
	in the simple energy shift indicated in Eqs.~(\ref{eq:ModSigma-}),(\ref{eq:ModSigma+}).}
\label{fig:SigmaRPol}
\end{figure}


We have calculated the full Keldysh self-energy corresponding to  $\hH_2$ to order $g^2$, without any further approximation.  We have also done this calculation in the case where $G$ is not so large, such that one should work in the original basis of photons and phonons.  Results of these full calculations will be presented elsewhere.  Here, like in the main text, we will focus on the resonant-interaction regime described in the main text, where the condition $E_+ = 2 E_-$ enhances certain scattering processes.  In this regime, the dominant self-energy processes for $+$ and $-$ polaritons are depicted in Fig.~\ref{fig:SigmaRPol}.  Using the standard rules of the Keldysh technique \cite{Kamenev09}, these diagrams correspond to:
\begin{eqnarray}
        \Sigma _{--}^{R}\left[ \omega \right]  &=&%
        2i\widetilde{g}^{2}\int \frac{d\omega^{\prime}}{2\pi}\left( \mathcal{G}_{--}^{K}\left[ \omega^{\prime}\right] \mathcal{G}_{++}^{R}\left[ \omega^{\prime}+\omega \right]%
        +\mathcal{G}_{--}^{A}\left[ \omega ^{\prime }-\omega \right] \mathcal{G}_{++}^{K}\left[ \omega ^{\prime }\right] \right)%
        =4\widetilde{g}^{2}\frac{\overline{n}_{-}-\overline{n}_{+}}{\omega -\left( E_{+}-E_{-}\right) +i\frac{\kappa _{-}+\kappa _{+}}{2}}, \\
        \Sigma _{++}^{R}\left[ \omega \right]  &=&2i\widetilde{g}^{2}\int \frac{d\omega ^{\prime }}{2\pi }\mathcal{G}_{--}^{K}\left[ \omega ^{\prime }\right] \mathcal{G}^{R}\left[ \omega - \omega ^{\prime }\right] %
        =2\widetilde{g}^{2} \frac{\left( 1+2\overline{n}_{-}\right) }{\omega -2E_{-}+i\kappa _{-}}.
    \end{eqnarray}

Note that because of the resonance condition, these self-energies scale as $g^2 / \kappa$, whereas all other self-energy diagrams are suppressed by an additional small parameter $\kappa / \omegam$.  They can thus be neglected in the limit $\kappa / \omegam \rightarrow 0$.  For a small but realistic value of  $\kappa / \omegam$, the small-energy shifts associated with the non-resonant interaction terms in $\hH_2$ can shift the condition needed for resonance.  To describe these small shifts (which can be important for realistic parameters), it is sufficient to use standard second-order perturbation theory to treat the non-resonant terms.  This is conveniently done via a Schrieffer-Wolff transformation, where a unitary transformation is performed to eliminate the non-resonant terms to leading order in the Hamiltonian.  The procedure generates new terms however at second order in $g$.  Keeping only such terms which do not change the total number of polaritons, we obtain the general form given in the Hamiltonian $\hH_{NR}$ (Eq.~(8) of the main text).  Note that in this equation, 
the energy shifts $\delta_\sigma$ include both the shifts arising from the non-resonant terms, as well as the energy shifts $\epsilon_\sigma$ coming from the normal-ordering procedure.  Both such terms scale as $g^2 / \omegam$.  

Including the energy shifts associated with the non-resonant processes (as described by $\hH_{NR}$ in Eq.~(8) of the main text), the self-energies above are modified to:
\begin{eqnarray}
        \Sigma _{--}^{R}\left[ \omega \right]  &=&%
        4\widetilde{g}^{2}\frac{\overline{n}_{-}-\overline{n}_{+}}{\omega -  \tilde{E}_{+-} +i\frac{\kappa _{-}+\kappa _{+}}{2}}, 
        \,\,\,\,\,\,\,\,\,
        \tilde{E}_{+-} = E_+ - E_- + \delta_+ - \delta_-  + U_{+-} \left( \bar{n}_{-} - \bar{n}_{+} \right), 
        \label{eq:ModSigma-}
        \\
        \Sigma _{++}^{R}\left[ \omega \right]  &=&
        2\widetilde{g}^{2} \frac{\left( 1+2\overline{n}_{-}\right) }{\omega -\tilde{E}_{--}+i\kappa _{-}},
        \,\,\,\,\,\,\,\,\,
        \tilde{E}_{--} = 2 \left(E_{-} + \delta_{-} \right) + 2 U_{--}\left(1 + 2 \bar{n}_{-} \right).
        \label{eq:ModSigma+}
    \end{eqnarray}
The corrections due to the Kerr-type interaction constants $U_{+-}$ and $U_{--}$ can be obtained by including ladder diagrams in the self-energies, as shown in Fig.~\ref{fig:SigmaRPol}b.

\section{OMIT reflection coefficient}

In an OMIT style-experiment, in addition to the main driving laser (which gives rise to the many-photon interaction $G$), a second weak drive tone (the ``probe") is applied at a frequency $\omega_{\rm pr}$ to the cavity.
This driving is described by a term in the Hamiltonian:
\begin{eqnarray}
	\hH_{\rm pr} = -i \sqrt{\kappa_{\rm cp}} \left( \hd^\dag \bar{d}_{\rm in,pr} e^{-i \tilde{\omega}_{\rm pr} t} + h.c. \right)
	\label{eq:Hprobe}
\end{eqnarray}
where we work in the interaction picture determined by the main drive laser frequency, and hence $\tilde{\omega}_{\rm pr} = \omega_{\rm pr} - \omega_L$.  $\kappa_{\rm cp}$ parameterizes the coupling of the drive port to the cavity:  for a one-port cavity, the total cavity damping rate $\kappa = \kappa_{\rm cp} + \kappa_{\rm int}$, where $\kappa_{\rm int}$ describes internal cavity losses.

As the amplitude $\bar{d}_{\rm in}$ probe drive is weak, we can use standard linear response theory to describe its effects.  The Kubo formula thus tells us that to first order in $\bar{d}_{\rm in, pr}$, the change in the 
cavity amplitude will be given by:
\begin{eqnarray}
	\delta \left\langle \hat{d}(t) \right  \rangle
		&= &  -i \sqrt{\kappa_{\rm cp}}  \int_{-\infty}^{\infty} dt'  \left(
				\bar{d}_{\rm in, pr} G^R_{dd}(t-t') e^{-i \tilde{\omega_{\rm pr}} t'} -
				\bar{d}_{\rm in, pr}^* G^R_{d\bar{d}}(t-t') e^{i \tilde{\omega_{\rm pr}} t'} 
			\right)	\\
		& = &
			 -i \sqrt{\kappa_{\rm cp}}
			 \left(
			 	   e^{-i \tilde{\omega}_{\rm pr} t} \bar{d}_{\rm in, pr}   G^R_{dd}[ \tilde{\omega}_{\rm pr} ] -
			 	  e^{i \tilde{\omega}_{\rm pr} t}    \bar{d}_{\rm in, pr}^*  G^R_{d \bar{d}}[ -\tilde{\omega}_{\rm pr} ] 
			\right)  .
\end{eqnarray}
Here, $G^R_{dd}[\omega]$ is the retarded Green function of the cavity as defined above, calculated to zeroth order in $\hH_{\rm pr}$, but including the effects of the nonlinear interaction $g$.  $G^R_{d \bar{d}}[\omega]$ is the corresponding anomalous Green function defined as:
\begin{eqnarray}
        G_{d \bar{d} }^{R}\left[ \omega \right]  &\equiv &
        -i\int dt\theta(t) \left\langle \left[ \hat{d}(t),\hat{d}(0)\right] \right\rangle e^{i\omega t},  \label{eq:GAnom}
\end{eqnarray}

Now, the standard input-output relation between input, output and cavity fields is \cite{ClerkRMP}:
\begin{eqnarray}
	\hat{d}_{\rm out}(t) = \hat{d}_{\rm in}(t) + \sqrt{\kappa_{\rm cp}} \hat{d}(t)
\end{eqnarray}
Taking the average value of this equation, and defining the elastic amplitude reflection coefficient $r[\omega_{\rm pr}]$ as the amplitude of $\langle \hat{d}_{\rm out}(t) \rangle$ at the probe frequency divided by $\bar{d}_{\rm in}$, we obtain
\begin{equation}
	r[\omega_{\rm pr}]  =   1 - i  \kappa_{\textrm{cp}} G^R_{dd}[\omega_{\rm pr}],
\end{equation}
as given in the main text.

%



\section{Master equation simulation}

Starting from the system-bath Hamiltonians written in the polariton basis (Eqs.~(\ref{eq:HCavBath}) and (\ref{eq:HMechBath})), one can trace over the dissipative baths
and derive a master equation for the reduced density matrix $\hrho$ describing the polaritons using standard Born-Markov approximations~\cite{Gardiner00}.
One obtains:
\begin{equation}
	\partial_t \hrho = -i \left[ \hH_1 + \hH_2, \hrho \right] + \mathcal{L} \hrho
	\label{eq:Master}
\end{equation}
where the coherent system Hamiltonian $\hH_1 + \hH_2$ is written without any approximation, and the super-operator $\mathcal{L}$ describes the effects of the dissipative baths
via standard Lindblad terms,
%
%
%
%
\begin{equation}
\mathcal{L}=\kappa_-(1+\bar{n}_-)L[\hcm_-]+\kappa_-\bar{n}_-L[\hcm_-^\dag]+\kappa_+(1+\bar{n}_+)L[\hcm_+]+\kappa_+\bar{n}_+L[\hcm_+^\dag],
\label{lindbladian}
\end{equation}
with
\begin{equation}
L[\hcm]\cdot=\hcm\cdot\hcm^\dag-\frac{1}{2}\{\hcm^\dag\hcm,\cdot\}.
\end{equation}
Note that this master equation corresponds to each polariton seeing independent thermal baths, in direct analogy to the form of the quantum Langevin equations in Eq.~(\ref{eq:Langevins}).

While Eq.~(\ref{eq:Master}) is not a convenient starting point for deriving analytic results, it does allow us to numerically study the system without having to assume a small value of $g$.  Using
Eq.~(\ref{eq:Master}) and the quantum regression theorem \cite{Gardiner00}, we have numerically calculated the cavity density of states $\rho_d[\omega] = - \textrm{Im } G^R_{dd}[\omega] / \pi$, 
finding good agreement with our analytic perturbative results even for $g$ as large as $\kappa$ (see Fig.~1 in main text).  Note that to find agreement with these numerical results, it was crucial to include in the analytic theory the corrections associated with the non-resonant interaction processes, $\hH_{NR}$
 (c.f.~Eqs.(\ref{eq:ModSigma-}) and (\ref{eq:ModSigma+})).  

Finally, we have also numerically studied our system using a more conventional master equation, in which the dissipation is described by Lindblad superoperators which act in the photon and phonon basis, i.e.~replace $\mathcal{L}$ in Eq.~(\ref{eq:Master}) with $\mathcal{L}_0$, where:
\begin{equation}
	\mathcal{L}_0 = 
		\kappa L[\hd]  + \gamma(1+\bar{n}_M)L[\hb] + \gamma \bar{n}_ML[\hb^\dag],
\label{lindbladian}
\end{equation}
and $\bar{n}_M$ is a Bose-Einstein distribution evaluated at the mechanical frequency $\omegam$ and mechanical bath temperature.
For the parameters studied in the paper, this conventional master equation yields results very similar to those obtained from Eq.~(\ref{eq:Master}).

%



\section{$g_2$ correlation functions}

As discussed in the main text, the resonant two-particle interaction between $-$ polaritons can lead to non-classical values of the $g_2$ correlation functions.  In Fig.~3 of the main text, we have used a numerical solution of our master equation and the quantum regression theorem to calculate $g_2$ functions of both photons and $-$ polaritons, taking $g = \kappa$, $\Delta = - \omegam$ and tuning $G$ to $G_{\rm res} = 0.3 \omegam$ to ensure the resonance condition $E_+ = 2 E_-$.  For simplicity, we also take the large $\omegam$ limit (i.e.~ $\omegam / \kappa \rightarrow \infty$, $g / \omegam \rightarrow 0$), so that corrections due to non-resonant interaction terms in $\hH_2$ can be neglected.  
We have also included a weak probe drive on the cavity at a frequency $\omega_{\rm pr}$ close to $E_{-}$ by including a term of the form
$\hH_{\rm pr}$ (c.f.~Eq.(\ref{eq:Hprobe})).  When writing this driving term in terms of polarities, we will have terms that effectively drive both $-$ and $+$ 
polaritons.  However, in the large $\omegam$ limit and for $\omega_{\rm pr} \simeq E_-$, the direct driving term on the $+$ polaritons will be strongly off-resonance and can be neglected.  The probe field driving Hamiltonian thus reduces to:
\begin{eqnarray}
	\hH_{\rm pr} = 
		\epsilon e^{-i \omega_{\rm pr} t} \hat{c}_{\rm -}^\dag + h.c.
		\label{eq:Hpr2}
\end{eqnarray}

Eq.~(15) of the main text and the dashed curve in Fig.~3 are the results of a simple analytic theory which accurately describes the $-$ polariton $g_2$ function in the limit of 
an extremely weak probe drive amplitude $\epsilon$.  For a weak drive (and only keeping the resonant interaction process in $\hH_2$), we can restrict attention to the four lowest eigenstates of the coherent Hamiltonian.  These are:
\begin{align}
\ket{\psi_0}&=\ket{0,0},
&E_0&=0,\\
\ket{\psi_1}&=\ket{1,0},
&E_1&=E_-,\\
\ket{\psi_2}&=-\sin\theta\ket{0,1}+\cos\theta\ket{2,0},
&E_2&=\tfrac{1}{2}E_++E_--\sqrt{(\tfrac{1}{2}E_+-E_-)^2+2\tilde{g}^2},\\
\ket{\psi_3}&=\cos\theta\ket{0,1}+\sin\theta\ket{2,0},
&E_3&=\tfrac{1}{2}E_++E_-+\sqrt{(\tfrac{1}{2}E_+-E_-)^2+2\tilde{g}^2},
\end{align}
where $\ket{n_-,n_+}$ denotes a Fock state having $n_-$ $-$ polaritons and $n_+$ $+$ polaritons, and $\tan2\theta=2\sqrt{2}\tilde{g}/(E_+-2E_-)$.
The system is driven by adding the probe drive Hamiltonian $\hH_{\rm pr}$ given in Eq.~(\ref{eq:Hpr2}).
Including the drive, the wave function is written in terms of the above states as:
\begin{equation}
\ket{\psi}=A_{00}\,\ket{0,0}+A_{10}\,\ket{1,0}+A_{20}\,\ket{2,0}+A_{01}\,\ket{0,1}.
\end{equation}
We take into account the dissipation with the anti-Hermitian Hamiltonian 
$\hH_\mathrm{damping}=-\tfrac{1}{2}i(\kappa_-\hcm_-^\dag \hcm_-+\kappa_+\hcm_+^\dag \hcm_+)$.
In the rotating frame corresponding to the unitary operator $\Exp{i(\hcm_-^\dag\hcm_-+2\hcm_+^\dag\hcm_+)\omega_{\rm pr} t}$, the Hamiltonian reads
\begin{equation}
H_\mathrm{RWA}=(\Delta_{-}-\tfrac{i}{2}\kappa_-)\hcm_-^\dag \hcm_-+(\Delta_+-\tfrac{i}{2}\kappa_+)\hcm_+^\dag \hcm_++\tilde{g}(\hcm_+^\dag \hcm_-\hcm_-+\hcm_+\hcm_-^\dag \hcm_-^\dag)+\epsilon(\hcm_-^\dag+ \hcm_-),
\end{equation}
with the detuning $\Delta_- = E_- - \omega_{\rm pr}$ and $\Delta_+ = E_+-2\omega_{\rm pr}=2\Delta_-$ at the resonance $E_+=2E_-$.
The dynamics is given by $\partial_t\ket{\psi}=-i \hH\ket{\psi}$.
Ignoring other states than those involved in $\ket{\psi}$, the steady state number and correlation
$g_{2-}=\moy{\hcm_-^\dag \hcm_-^\dag \hcm_- \hcm_-}/\moy{\hcm_-^\dag \hcm_-}^2$
are
\begin{align}
\moy{\hcm_-^\dag \hcm_-}&=\frac{\epsilon^2}{\Delta_-^2+(\kappa_-/2)^2},\\
g_{2-}&=\frac{[\Delta_-^2+(\kappa_-/2)^2][4\Delta_-^2+(\kappa_-/2)^2]}{[2\Delta_-^2-\tilde{g}^2-(\kappa_-/2)^2]^2+9\Delta_-^2(\kappa_-/2)^2},
\end{align}
where we approximate $A_{00}=1$, take $\kappa_+=\kappa_-$ and keep the leading terms in $\epsilon$.
The minimum value of $g_{2-}$ is found at $\omega_{\rm pr}=E_-$ with 
$g_{2-}=\kappa_-^4/(4\tilde{g}^2+\kappa_-^2)^2$
and the correlation tends toward unity for infinite detuning.

We comment briefly on the unusual double-peak resonance exhibited by $g_{2b}[\omega_{\rm pr}]$ in Fig.~3 of the main text.  This structure is also present in the behaviour of the average phonon number versus probe drive but not in the behaviour of the average photon number, see Fig.~\ref{fig:Populations}.  This difference is a higher-order effect of the probe drive-field that would not be captured in linear-response (i.e.~contributions to $\langle \hat{b}^\dagger \hat{b} \rangle$ beyond order $\epsilon^2$).  At higher orders, the
probe field causes non-zero correlations between $+$ and $-$ polaritons, e.g.~$\langle \hat{c}^\dagger_+ \hat{c}_- \rangle \neq 0$.  It follows from the normal-mode transformation of Eq.~(\ref{eq:UTransform}) that such averages contribute to the photon and phonon populations with opposite sign; this ultimately explains the strong difference in the behaviour of photon and phonon number versus 
$\omega_{\rm pr}$.


\begin{figure}[t]
	\begin{center}
	\includegraphics[width= 0.45\columnwidth]{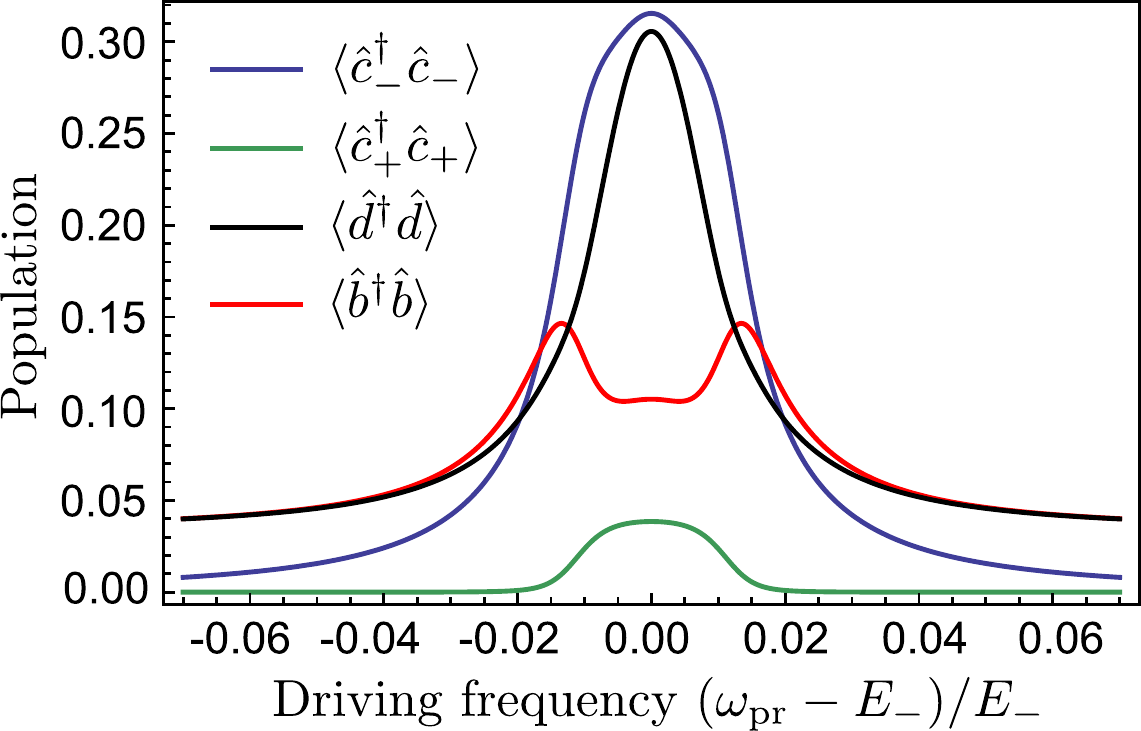}
	\end{center}
	\vspace{-0.5cm} 
	\caption {Plot of average polariton, photon and phonon numbers versus probe frequency $\omega_{\rm pr}$, for parameters identical to those used in Fig.~3
	of the main text.  We have taken a drive strength of $\epsilon = 0.2 \kappa$, where $\epsilon$ is defined in Eq.~(\ref{eq:Hpr2}).  The double-peak
	structure in the phonon population is due to higher-order effects of the probe drive which induce correlations between $+$ and $-$ polarities.	
	}
\label{fig:Populations}
\end{figure}

\bibliographystyle{apsrev}


\end{widetext}

\end{document}